\documentclass[preprint]{elsarticle}

\usepackage{graphicx}
\usepackage{amsmath}

\begin{document}

\title{The ORGAN Experiment: An axion haloscope above 15 GHz}

\author[equs]{Ben T. McAllister}
\ead{ben.mcallister@uwa.edu.au}
\author[equs]{Graeme Flower}
\author[phys]{Eugene N. Ivanov}
\author[equs]{Maxim Goryachev}
\author[equs]{Jeremy Bourhill}
\author[equs]{Michael E. Tobar}
\ead{michael.tobar@uwa.edu.au}
\address[equs]{ARC Centre of Excellence for Engineered Quantum Systems, School of Physics, The University of Western Australia, Crawley 6009, Australia}
\address[phys]{School of Physics, The University of Western Australia, Crawley 6009, Australia}
\date{\today}

\begin{abstract}
We present first results and future plans for the {\bfseries{O}}scillating {\bfseries{R}}esonant {\bfseries{G}}roup {\bfseries{A}}xio{\bfseries{N}} (ORGAN) experiment, a microwave cavity axion haloscope situated in Perth, Western Australia designed to probe for high mass axions motivated by several theoretical models. The first stage focuses around 26.6 GHz in order to directly test a claimed result, which suggests axions exist at the corresponding mass of $110~\mu$eV. Later stages will move to a wider scan range of 15-50 GHz ($60-210~\mu $eV). We present the results of the pathfinding run, which sets a limit on $g_{a\gamma\gamma}$ of $2.02\times 10^{-12} $eV$^{-1}$ at 26.531 GHz, or 110~$\mu$eV, in a span of 2.5 neV (shaped by the Lorentzian resonance) with $90 \%$ confidence. Furthermore, we outline the current design and future strategies to eventually attain the sensitivity to search for well known axion models over the wider mass range. 
\end{abstract}

\begin{keyword}
axions \sep dark matter \sep haloscope \sep ORGAN 
\PACS 14.80.Va \sep 95.35.+d
\end{keyword}
\maketitle

\section{Introduction}
\subsection{Axions and detection experiments}
The nature of dark matter is one of the greatest questions plaguing physics today. Many have suggested that dark matter is composed of weakly-interacting particles. One such class of particles of great interest are so called weakly-interacting slim particles, or WISPs~\cite{wisps}. The most well known and theoretically motivated WISPs are axions (or ``axion like particles", ALPs). Axions were first proposed in 1977 as a consequence of an elegant solution to the strong CP problem in QCD~\cite{PQ1977}. Subsequently it was proposed that axions might compose dark matter as they should have the desired properties, specifically, a mass and a weak coupling to regular matter~\cite{Sikivie1983b}.

The first direct, laboratory search technique for these particles was proposed in 1983 and is known as the haloscope, so named as it searches for axions in the galactic halo~\cite{Sikivie83haloscope,Sikivie1985}. The haloscope is a technique for detecting axions via their coupling to photons. In what is known as the inverse Primakoff effect an axion will convert into a photon when another photon is present. In most haloscopes this second photon is virtual, provided by an external static magnetic field. As a result the frequency of the generated real photon corresponds directly to the mass of the axion. There are a number of axion haloscope searches currently underway, most notably the Axion Dark Matter Experiment (ADMX), the first and most mature of such experiments~\cite{ADMXaxions2010,ADMX2011}.

One of the largest problems facing axion haloscopes is the fact that the mass of the axion is unknown (other than some broad cosmological limits~\cite{Sikivie1983,Preskill1983}), meaning that the frequency of the generated photons is also unknown. This, combined with the fact that the strength of the axion coupling to photons is unknown, creates a large parameter space for searching, which requires a number of experiments to probe different axion mass ranges. The most critical parameter in axion searches is the Peccei-Quinn symmetry breaking scale, $f_a$, which determines both the axion mass and the strength of its coupling to photons according to
\begin{equation*}
	\text{m}_a=\frac{z^{1/2}}{1+z}~\frac{f_{\pi}m_{\pi}}{f_a}~~~~
	\text{g}_{a\gamma\gamma}=\frac{\text{g}_{\gamma}\alpha}{f_{a}\pi}.
\end{equation*}
Here $\text{m}_a$ is the axion mass, $z$ is the ratio of up and down quark masses, $\frac{m_u}{m_d}~\approx$ 0.56, $f_\pi$ is the pion decay constant $\approx$ 93 MeV, $m_\pi$ is the neutral pion mass $\approx$ 135 MeV, $\text{g}_\gamma$ is an axion-model dependent parameter of order 1, and $\alpha$ is the fine structure constant~\cite{K79,Kim2010,DFS81,SVZ80,Dine1983}. $f_a$ is the parameter that haloscopes ultimately wish to constrain. A recent and highly promising theoretical model, known as SMASH, points to high mass ($\sim$100 $\mu$eV, or $\sim$24 GHz) axions, with corresponding frequencies in the microwave and millimetre wave ranges~\cite{SMASH}. Furthermore, a recent paper (hereafter referred to as the Beck result) claims that an anomalous effect observed in Josephson junctions can be explained due to an axion with a mass of around 110 $\mu$eV (26.6 GHz) entering the junction~\cite{Beck2013,Beck2014}. These factors combine to motivate direct axion searches above 12.5 GHz, particularly in the range near 25 GHz, and specifically around 26.6 GHz. However, due to a host of technical limitations~\cite{darin2001} and practical considerations, most haloscopes operate at frequencies well below 10 GHz, and this highly promising region of the axion parameter space remains largely unprobed. There is currently some interest in developing axion searches in mass ranges lower than those traditionally searched~\cite{Sikivie2014a,McAllister:2016fux,ABRACADABRA}, however, the move towards high frequencies and high axion masses is a more common contemporary goal of axion haloscopes~\cite{MADMAX,ADMXHF2014,YaleAxion,CULTASK,Jaeckel2013,SpringsPaper}.

The ORGAN experiment has performed a static (non frequency tunable) pathfinding run that allows very narrow limits to be placed on axion-photon coupling near 26.6 GHz, and development of later stages of the experiment, which will scan a large area of this highly promising mass range, is currently underway.
\subsection{Haloscopes}
A haloscope is an axion detection technique in which a resonant cavity is embedded in a strong (typically) static, uniform magnetic field. If axions are present in the cavity due to an abundance in the galactic halo dark matter, a small number will convert into real photons with a frequency corresponding to the axion mass. If the central frequency of the resonant cavity is tuned to overlap with this frequency, the signal will be resonantly enhanced, at which point it can be read out via low noise electronics, and ideally detected above the thermal noise of the system. The expected signal power in a haloscope is given by~\cite{Daw:1998jm}

\begin{equation}
\text{P}_a\propto\text{g}_{a\gamma\gamma}^2B^2CVQ_L\frac{\rho_a}{\text{m}_a}.
\label{eq:Paxion}
\end{equation}
Here $B$ is the field strength of the external magnetic field, $C$ is a mode dependent form factor of order 1~\cite{McAllisterFormFactor}, which represents the degree of overlap between the cavity mode electromagnetic field and the electromagnetic field induced due to axion photon conversion (it is an integral of the dot product of these two fields), $V$ is the volume of the detecting cavity, $Q_L$ is the loaded cavity quality factor (provided it is lower than the expected axion signal quality factor $\sim10^6$), and $\rho_a$ is the local axion dark matter density. The signal powers in axion haloscopes are extremely weak, even ADMX, which operates with a very large volume and high magnetic field system expects signal powers on the order of $10^{-22}$ W. The challenges faced in the move to high frequency haloscopes are evident in equation~\ref{eq:Paxion}. The volume of detecting cavities scales inversely with frequency, the surface resistance of materials increases with frequency, which impacts quality factors, and there is a $\frac{1}{m_a}$ term in the haloscope power equation. To further complicate this, the quantum noise limit of amplifiers increases with frequency, making it more difficult to resolve a signal above the noise of the amplifier. A common suggestion to alleviate the volume concern is to utilize a large number of small cavities power combined, however, this presents a series of immense practical challenges in an experiment. The scanning rate of a haloscope is given by~\cite{Stern:2015kzo}

\begin{equation}
\frac{df}{dt}\propto\frac{1}{\textit{SNR}_{goal}^2}\frac{\text{g}_{a\gamma\gamma}^4B^4C^2V^2\rho_a^2Q_LQ_a}{m_a^2(k_BT_n)^2},
\end{equation}
where $\textit{SNR}_{goal}$ is the desired signal-to-noise ratio of the search, $Q_a$ is the axion signal quality factor, and $T_n$ is the effective noise temperature of the first stage amplifier, with later amplifier contributions suppressed by the gain of this amplifier. This is the quantity that must be maximized in design of an experiment, for which $C^2V^2G$ can be viewed as a figure of merit for resonator design as all other parameters either depend on the properties of the axion, the external magnetic field or the first stage readout amplifier. $G$ is the mode geometry factor given by

\begin{equation}
G=\frac{\omega\mu_0\int{|\vec{H}|^2dV}}{\int{|\vec{H}|^2dS}}.
\end{equation}
Where $\vec{H}$ is the cavity mode magnetic field, rather than the external magnetic field. $G$ is directly proportional to the mode quality factor according to
\begin{equation}
Q=\frac{G}{R_s},
\end{equation}
where $\omega$ is the resonant frequency, $\mu_0$ is the permeability of free space, $\vec{H}$ is the cavity magnetic field, and $R_s$ is surface resistance of the material.
\subsection{ORGAN}
The ORGAN experiment will be an Australia-wide collaboration of different nodes from the ARC Centre of Excellence for Engineered Quantum Systems (EQuS), hosted at the University of Western Australia, and will search for axions in the range 15-50 GHz. The experiment will employ a 14 T superconducting magnet, and a variety of thin, long resonant cavities of different dimensions, similar in arrangement to a pipe organ. The initial stages of the experiment will utilize traditional HEMT-based amplification as the EQuS nodes develop quantum-limited amplification chains based on Josephson parametric amplifiers. The project has been in a development stage for the past few years, and has recently undertaken its pathfinding run, using a stationary frequency cavity mode, and a series of HEMT-based amplifiers at 4 K, inside a 7 T magnet.
\section{Pathfinding run}
\begin{figure}[t!]
	\centering
	\includegraphics[width=0.5\columnwidth]{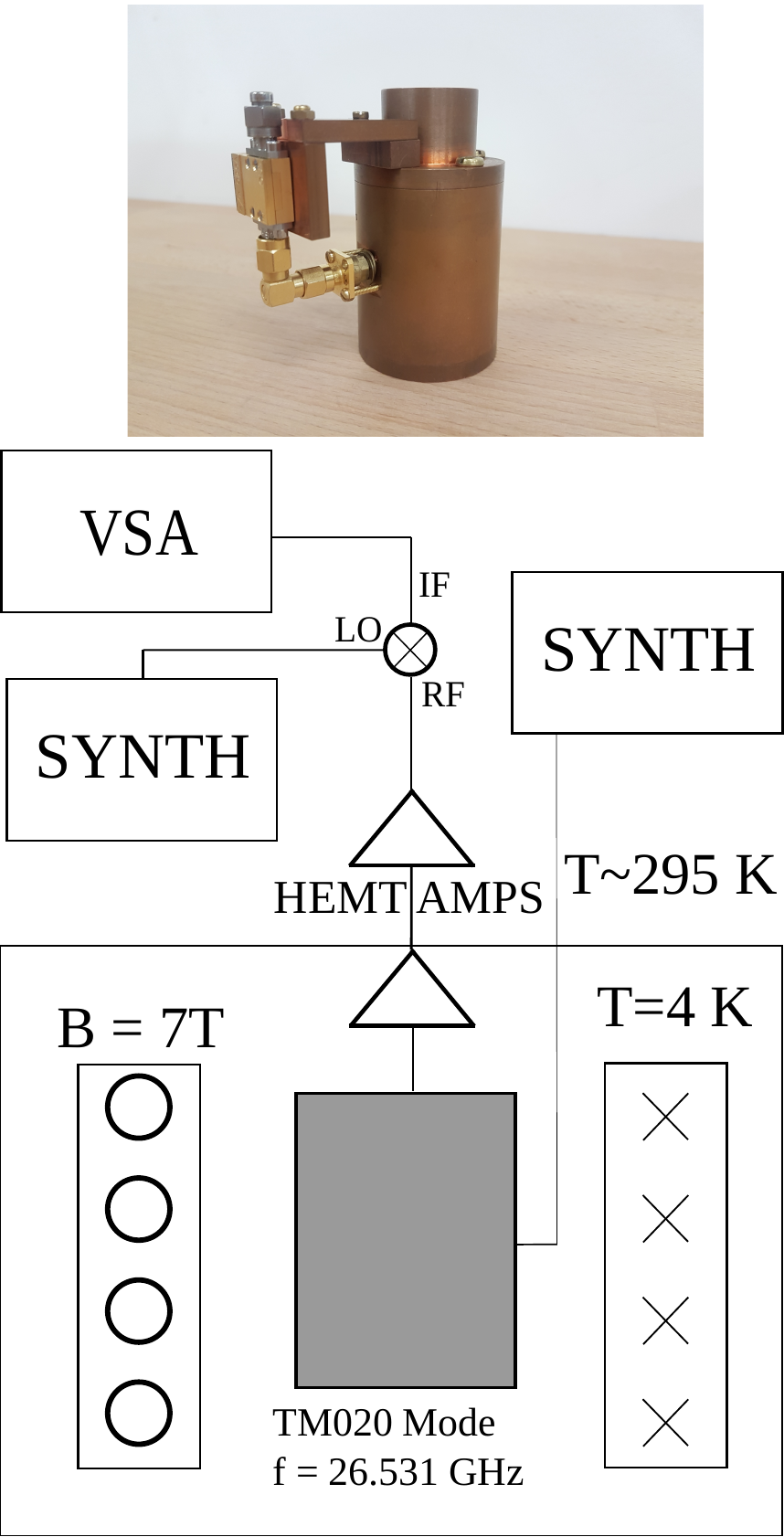}
	\caption{Top: the resonant cavity and first stage LNA employed. Bottom: Schematic of the pathfinder experiment. A single resonant cavity with a stationary $\text{TM}_{020}$ frequency of 26.531 GHz and a cryogenic HEMT amplifier with an effective noise temperature of 8 K were employed. The experiment collected data for 4 continuous days. The amplifiers were provided by Low Noise Factory and the mixer by Marki. The 26.531 GHz signal was mixed down to approximately 8 MHz in order to be sampled by a commercial VSA from Agilent.}
	\label{fig:pathfinderschematic}
\end{figure}
For the initial test run of the experiment it was decided to focus around 26.6 GHz, in order to prepare for a full-span, direct test of the Beck result, which suggests axions between 26.1 and 27.1 GHz (108 to 112 $\mu$eV). A small, oxygen-free copper resonant cavity with a $\text{TM}_{020}$ mode frequency of 26.531 GHz was manufactured. The $\text{TM}_{020}$ mode was chosen as it can be shown that $CV$ is constant between modes for a given frequency and length, and larger cavities are more practical to probe and characterize. Additionally, higher order modes tend to have higher geometry factors and thus higher quality factors at a given frequency~\cite{ORGANPatras2016}. This is illustrated in table~\ref{table:CVG}. Seeing as this was a stationary frequency test, mode crowding was not an issue. The specific resonant structures to be employed in future stages of the experiment are the subject of ongoing research and development. It is likely that they will employ some manner of dielectric structure based on~\cite{DielectricSupermode}, and that there will be a number of cavities operating simultaneously.

\begin{table}
	\centering
\begin{tabular}{|c|c|c|c|}
	\hline
Mode&Form Factor&Volume ($cm^3$)&Geometry Factor ($\Omega$)\\ \hline
$\text{TM}_{010}$&0.69&1.45&386.5\\ \hline
$\text{TM}_{020}$&0.13&7.78&744.6\\ \hline
$\text{TM}_{030}$&0.053&18.87&1244.3\\ \hline
\end{tabular}
	\caption{Electromagnetic Form Factor, Volume and Geometry Factor of a haloscope cavity for the first three axion sensitive modes at a fixed frequency and length.}
	\label{table:CVG}
\end{table}
\begin{figure}[t!]
	\centering
	\includegraphics[width=0.63\columnwidth]{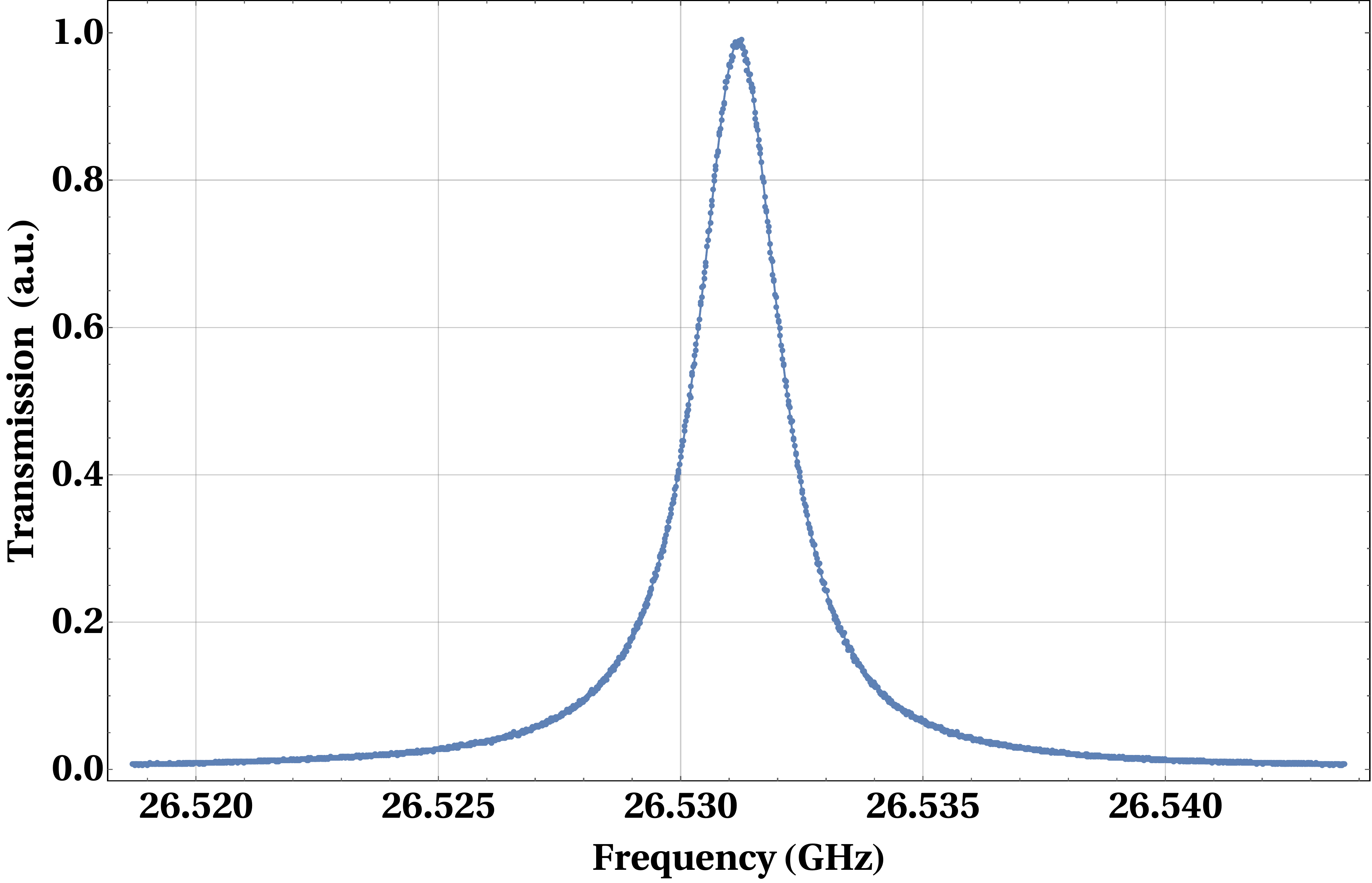}
	\includegraphics[width=0.68\columnwidth]{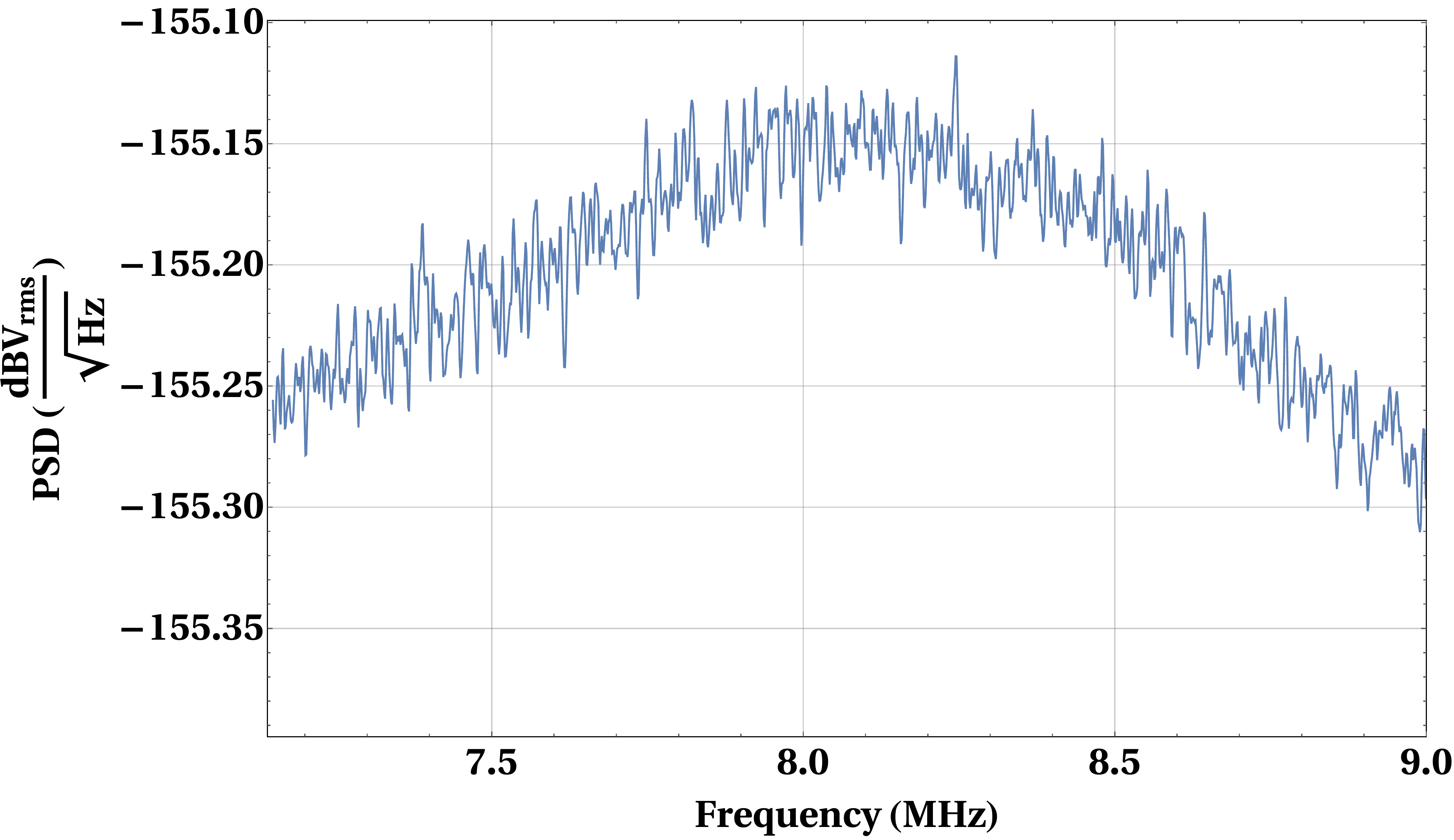}
	\caption{Two visualizations of the resonance, the first shows transmission through the cavity in arbitrary units as a function of frequency, whilst the second is a section of the power spectral density of voltage fluctuations as recorded by the FFT after the signal has been mixed down as per fig.1}
	\label{fig:resonance}
\end{figure}
The primary purpose of this pathfinding run was to test all of the equipment on hand and generate some experience with the requisite tools and processes. We verified that the magnet and cryostat could sustain long term, high-field operation, that our readout chain could resolve the cavity noise and that our digitization and data collection and analysis systems operated correctly. As a result of running the entire system simultaneously, we can place narrow limits on axions around 26.531 GHz below the sensitivity of CAST~\cite{CAST}. To our knowledge this is the first such experiment to ever be performed above 15 GHz.
A schematic of the pathfinding experiment is given in fig.~\ref{fig:pathfinderschematic}. This same schematic will apply more or less to the future stages of the experiment, with varied parameters and a varied number of cavities. The resonance employed is shown in fig.~\ref{fig:resonance} and had a loaded cryogenic quality factor with a critically coupled probe of $\sim$13,000. The magnet, cavity and cryogenic HEMT-based amplifier (a Low Noise Factory LNF-LNC15$\_$29B with an effective noise temperature of roughly 8 K) were held at 4 K. This was achieved by attaching a conducting rod to the 4 K plate of the dilution refrigerator which ran directly into the magnet bore and supported the cavity and amplifier, without being in thermal contact with the mixing chamber plate or the magnet itself. Another Low Noise Factory amplifier was attached at the room temperature output of the cryogenic readout chain, followed by a mixer to down convert the cavity spectrum to be centred at $\sim$ 8 MHz, where it was sampled by a commercial vector signal analyser and recorded to an external computer for analysis. Later stages will employ an FPGA based digitizer with customized software, in development by another EQuS node.

\begin{figure}[t]
	\centering
	\includegraphics[width=0.85\columnwidth]{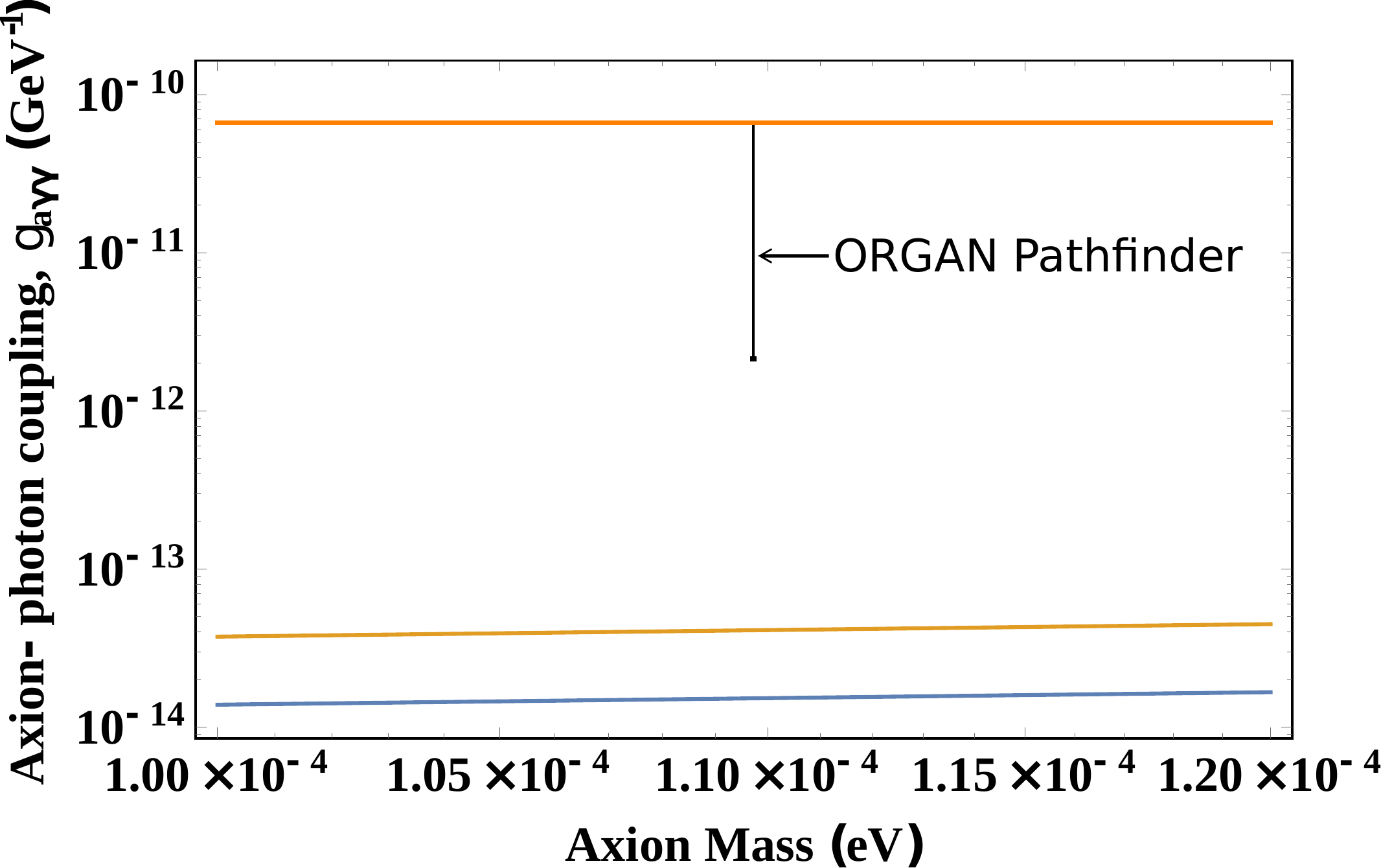}
	\caption{The narrow exclusion limits on axions for the stationary cavity pathfinding experiment. The axion KSVZ and DSFZ model bands, as well as exclusion limits from the CERN Axion Solar Telescope (CAST) are shown (orange).}
	\label{fig:narrowlimits}
\end{figure}
The experiment collected data for 4 continuous days. The data was then combined and linearly averaged, before a data analysis process similar to that undertaken in~\cite{Daw:1998jm} was performed: the power spectrum was computed with a bin width of 26.6 kHz, which is the linewidth of the axion signal assuming a central frequency of 26.6 GHz and a signal quality factor of $10^6$. At the end of the high field operation, approximately 12 hours of cold data was recorded with the magnet at 0 T, in order to exclude spurious systematic signals in the system. A 1.1~$\sigma$ cut was made on the averaged, background subtracted data, allowing us to exclude a signal of 3.4~$\sigma$ with 90\% confidence. When considering the relevant experimental parameters and employing equation.~\ref{eq:Paxion}, the resulting 90\% confidence exclusion limits on g$_{a\gamma\gamma}$ are shown in fig.~\ref{fig:narrowlimits}. Seeing as this is an untuned empty cavity, we use the theoretical value of $\sim$0.13 for the form factor, which is found both analytically and via finite element analysis. The physical temperature of the system was monitored closely via temperature sensors at different stages of the dilution refrigerator, and the noise temperature of the first stage cryogenic amplifier was measured separately on a commercial network analyser, and was further estimated via fitting to the thermal noise spectrum at the output of the system. The quality factor and probe coupling of the resonance were measured on earlier cool-downs, so that unloaded Qs could be obtained.
\section{Future Experiments}
The ORGAN experiment has recently received seven years of funding through the ARC Centre of Excellence for Engineered Quantum Systems. As such we outline plans to perform a series of long-term data collection experiments, each targeting a different region of the promising, unprobed axion parameter space. 
\subsection*{Stage I}
\begin{figure}[t]
	\centering
	\includegraphics[width=0.9\columnwidth]{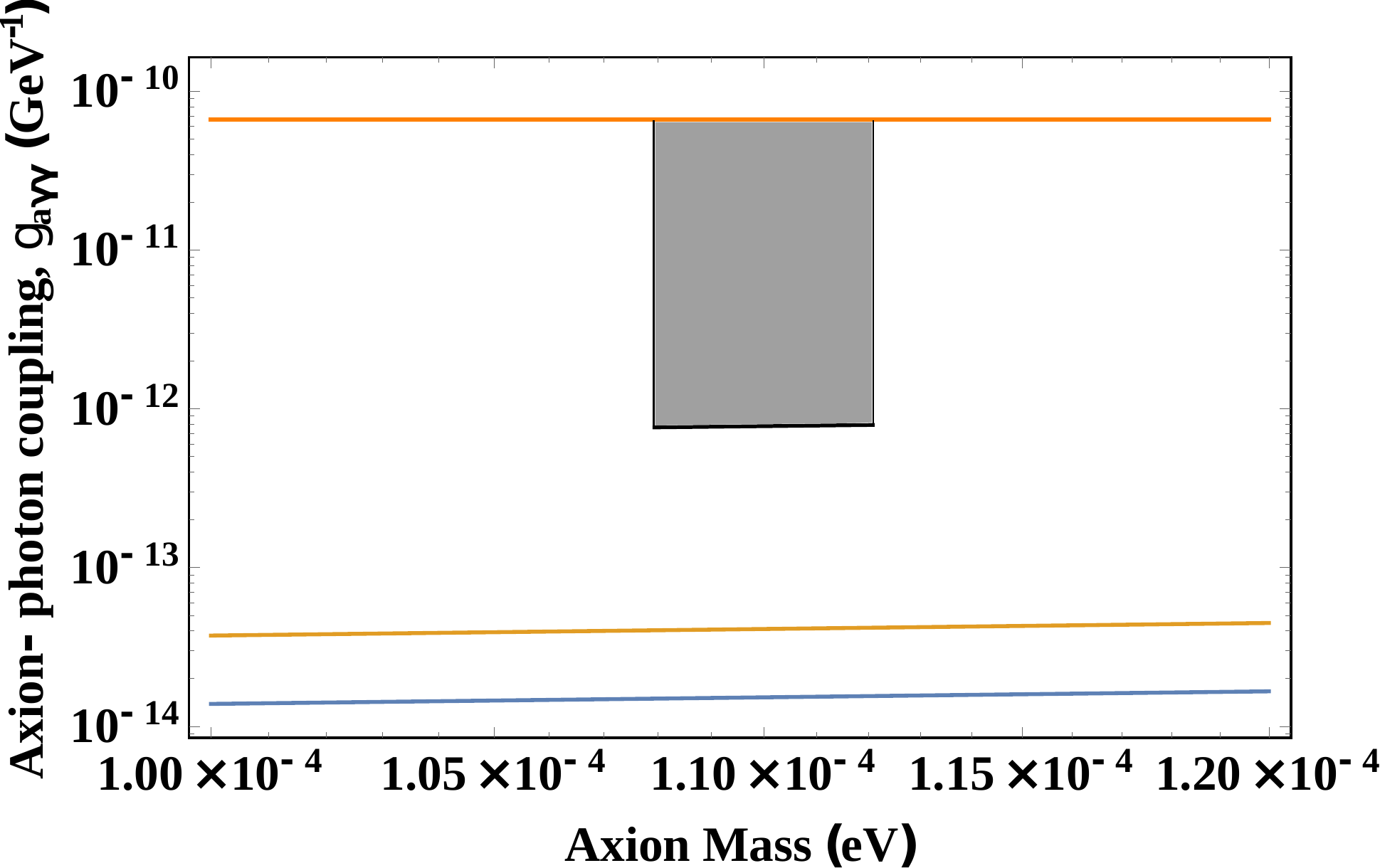}
	\caption{The projected exclusion limits for the first stage of the experiment, a targeted scan of the Beck result between 26.1 and 27.1 GHz with the best available equipment.}
	\label{fig:BeckTest}
\end{figure}

The first stage, currently planned to operate for the duration of 2018 will utilize two small cavities combined using cross-correlation~\cite{XSWisp}, embedded in a 14 T magnet at 30 mK, with traditional HEMT based amplification at 4 K. This experiment will focus on the frequency range 26.1 - 27.1 GHz, in order to undertake a direct test of the Beck result to the best sensitivity achievable with the technology available to us at the beginning of the tuning run. The resonator will utilize a dielectric tuning mechanism that is the subject of ongoing research, but will likely employ a dielectric supermode tuning mechanism as per~\cite{DielectricSupermode}. We anticipate mode quality factors on the order of 50,000 and a form factor on the order of 0.3 at 26.6 GHz. The projected sensitivity for this search is shown in fig.~\ref{fig:BeckTest}. Projected limits were generated by integrating the inverse of equation 2 over the frequency range and solving for $\text{g}_{a\gamma\gamma}$ inputting a fixed value of measurement time, with a signal to noise ratio of 5, assuming a dark matter density of $0.45$~$\frac{GeV}{cm^3}$~\cite{cdmdensity2014}.
\subsection*{Stage II}
During the operation of Stage I the ORGAN collaboration plans to prototype and develop for Stage II, a wider scan from 15-50 GHz to take place over the next 6 years, from 2019 - 2025. The scan will be undertaken in 5 GHz regions, Stage II-A through Stage II-G. Each region will receive approximately 10 months of data collection, during which time further research and development for the remaining future stages can be undertaken.
The aims of this research will be to develop resonators at the requisite frequencies with the best possible sensitivity based on a $C^2V^2G$ figure of merit, to develop quantum limited amplification for each frequency range, as well as a potential further magnet upgrade to 28 T. It is our goal to install quantum limited amplification at the beginning of Stage II-A at 15 GHz, and develop new amplification as we progress.
\begin{figure}[t!]
	\centering
	\includegraphics[width=\columnwidth]{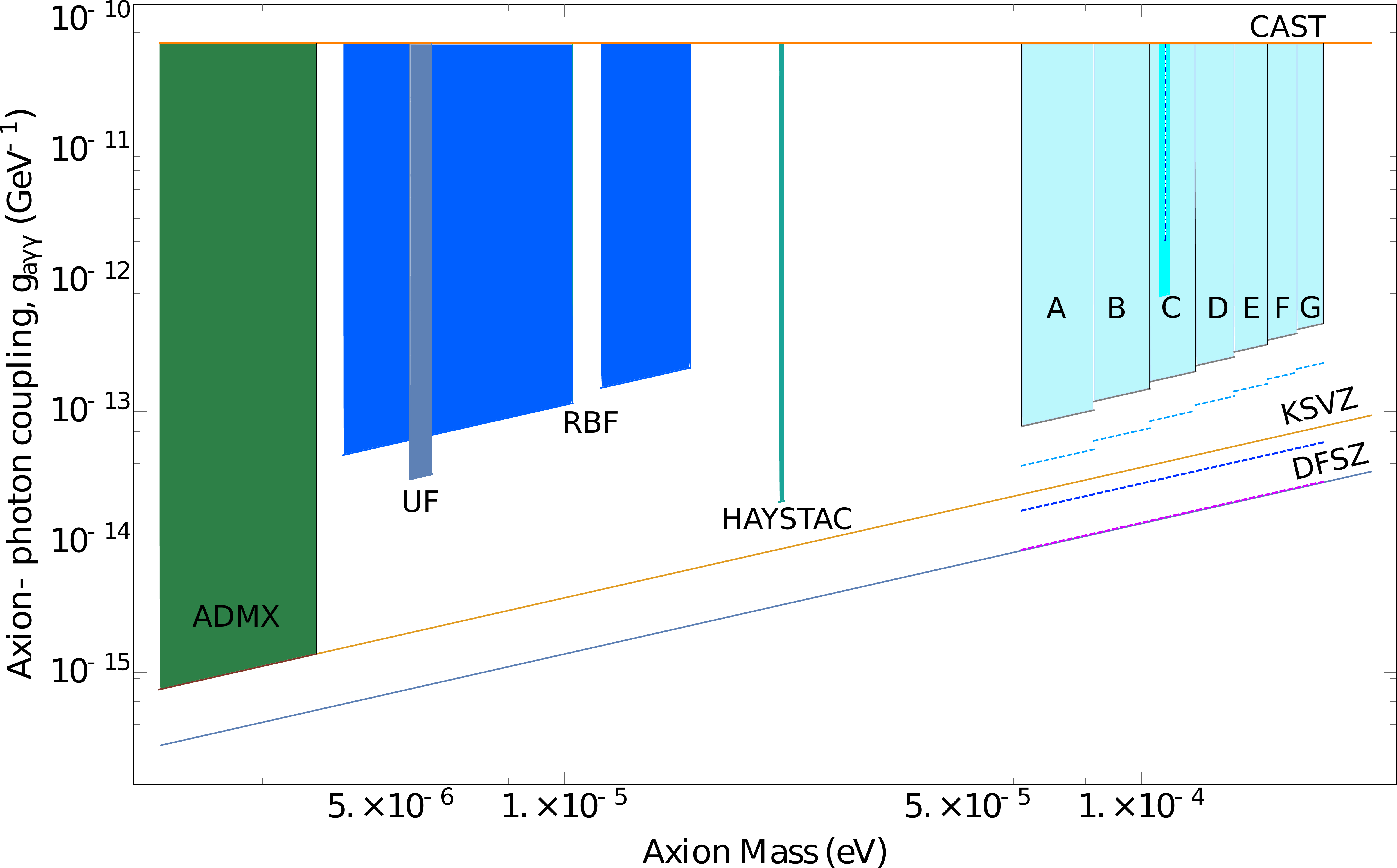}
	\caption{Strength of axion to two photon coupling as a function of axion mass with popular axion models outlined by the yellow and blue lines. Published exclusion limits from CAST (orange), ADMX~\cite{ADMXaxions2010,ADMX2011} (dark green), RBF~\cite{RBF1,RBF2} (blue), UF~\cite{UFaxions} (blue-gray), and HAYSTAC~\cite{YaleAxion} (aqua) are shown. The projected sensitivities for various ORGAN scenarios are displayed. The narrow cyan bar inside section C represents the first stage of the experiment, a targeted one year search in the region 26.1-27.1 GHz, assuming a 14 T magnet, but otherwise retaining the traditional HEMT-based amplification currently available. The dashed, narrower dark blue bar set within the cyan bar represents the limits from the path-finding experiment conducted at UWA with the existing infrastructure. A-G represent the various phases of the second stage of the experiment over the next six years of EQuS. The 15-50 GHz scan is broken into 5 GHz regions, assuming that the goal of quantum limited amplification is met over the course of the ORGAN experiment, with assistance from the EQuS collaboration. The dashed, light blue lines show the extension to these limits that could be reached with a larger 28 T magnet. The lower, dashed, darker blue line shows the projected sensitivity under the assumption that we can surpass the quantum noise limit and achieve added noise at the level of the thermal noise of the resonator and amplifier with the 14 T magnet, whereas the dashed purple line represents the same level of thermal noise for the larger 28 T magnet. If we meet our long term goals to surpass the quantum noise limit, and to work with an even stronger magnet it is within the realm of possibility to probe the lower end of the expected axion coupling range over the frequency range 15-50 GHz.}
	\label{fig:fullplot}
\end{figure}
Furthermore, collaborators at the Australian National University have experience developing squeezing techniques to beat the standard quantum limit of added noise for the Advanced LIGO experiment~\cite{ANULIGO}. The ORGAN collaboration will investigate implementing squeezed microwave states and quantum non-demolition measurements in order to surpass the quantum limit in the haloscope, and reach sensitivities inside the DFSZ to KSVZ model band. Such technologies and schemes have been proposed for use in low noise microwave experiments, and in axion detection experiments before~\cite{Nobel2012,AxionQND3,AxionQND1,AxionQND2}. The sensitivities for Stages II-A through II-G are shown in fig.~\ref{fig:fullplot}. The caption details the different scenarios associated with each set of exclusion limits, and experimental parameters are assumed to be similar to those expected in Stage I, with the assumption that at higher frequencies we will employ multiple resonators so that the effective volume utilization remains roughly constant. Improving on these parameters and thus further improving sensitivity will be a goal of ongoing research and development.
Each frequency range will require different cavity parameters and a different number of cavities. The planned 14 T magnet has a bore size of 65 mm and a length of 445 mm. As such, the cavities used for the lower end of the search will need to fill the radial size of the bore to reach the 15 GHz target (using a TM$_{030}$ mode cavity designed as per~\cite{DielectricSupermode}. As the central search frequency increases a larger number of cavities will be employed such that the effective volume utilization remains roughly constant. The exact structure of the resonators that will be employed for the later stages of the search is not finalized, but it is likely that 4-6 cavities will be utilized.
We have elected to begin our search frequency range at 15 GHz, as the region below this will be capably covered by other experiments such as those underway at ADMX, CULTASK, and HAYSTAC, and future experiments such as ABRACADABRA~\cite{ABRACADABRA}. Other planned experiments such as MADMAX~\cite{MADMAX} and ORPHEUS~\cite{Orpheus} aim to cover a similar mass range. As such, ORGAN will commence with the targeted search around 110~$\mu$eV, which will begin in 2018. Once this is complete, the experiment plans to cover the wider search range, but the multi-stage structure of the experiment will allow ORGAN to remain manoeuvrable, and capable of shifting focus to cover areas of the high mass parameter space that are not excluded, should one of these other high mass experiments successfully exclude some of the planned search space. Regions above 50 GHz are not as well motivated theoretically.
\section{Conclusion}
The ORGAN experiment has undertaken its pathfinding run, which constrains axion-photon coupling in a narrow span around 26.531 GHz. The best limit reached is $g_{a\gamma\gamma}>2.02\times 10^{-12} $eV$^{-1}$ with $90 \%$ confidence, or $\sim$50 times KSVZ at this frequency. Development of later stages of the experiment is underway.  If the goals outlined in the text are achieved, we will perform the first direct test for dark matter axions below the sensitivity of CAST in the highly promising region of 15-50 GHz utilizing a 14 T magnet and a series of small resonators. Through the ARC Centre of Excellence for Engineered Quantum Systems (EQuS) the experiment has funding for seven years. EQuS nodes will develop quantum limited amplification, whilst researching implementation of sub-quantum limited amplification, in addition to novel resonators. The first scanning stage of the experiment is planned to commence in 2018 and run for approximately a year. After this, 6 further years of searching are planned.\\
\\
This work was funded by Australian Research Council (ARC) grant no.CE110001013, as well as the Australian Government's Research Training Program and the Bruce and Betty Green foundation. The authors thank Stephen Parker for his significant contribution in the early stages of the project.
\bibliographystyle{elsarticle-num}

\begin{thebibliography}{10}
\expandafter\ifx\csname url\endcsname\relax
  \def\url#1{\texttt{#1}}\fi
\expandafter\ifx\csname urlprefix\endcsname\relax\def\urlprefix{URL }\fi
\expandafter\ifx\csname href\endcsname\relax
  \def\href#1#2{#2} \def\path#1{#1}\fi

\bibitem{wisps}
J.~Jaeckel, A.~Ringwald,
  \href{http://www.annualreviews.org/doi/abs/10.1146/annurev.nucl.012809.104433}{The
  low-energy frontier of particle physics}, Annual Review of Nuclear and
  Particle Science 60~(1) (2010) 405--437.
\newblock \href {http://dx.doi.org/10.1146/annurev.nucl.012809.104433}
  {\path{doi:10.1146/annurev.nucl.012809.104433}}.
\newline\urlprefix\url{http://www.annualreviews.org/doi/abs/10.1146/annurev.nucl.012809.104433}

\bibitem{PQ1977}
R.~D. Peccei, H.~R. Quinn,
  \href{http://link.aps.org/doi/10.1103/PhysRevLett.38.1440}{$\mathrm{CP}$},
  Phys. Rev. Lett. 38 (1977) 1440--1443.
\newblock \href {http://dx.doi.org/10.1103/PhysRevLett.38.1440}
  {\path{doi:10.1103/PhysRevLett.38.1440}}.
\newline\urlprefix\url{http://link.aps.org/doi/10.1103/PhysRevLett.38.1440}

\bibitem{Sikivie1983b}
J.~Ipser, P.~Sikivie,
  \href{http://link.aps.org/doi/10.1103/PhysRevLett.50.925}{Can galactic halos
  be made of axions?}, Phys. Rev. Lett. 50 (1983) 925--927.
\newblock \href {http://dx.doi.org/10.1103/PhysRevLett.50.925}
  {\path{doi:10.1103/PhysRevLett.50.925}}.
\newline\urlprefix\url{http://link.aps.org/doi/10.1103/PhysRevLett.50.925}

\bibitem{Sikivie83haloscope}
P.~Sikivie,
  \href{http://link.aps.org/doi/10.1103/PhysRevLett.51.1415}{Experimental tests
  of the "invisible" axion}, Phys. Rev. Lett. 51 (1983) 1415--1417.
\newblock \href {http://dx.doi.org/10.1103/PhysRevLett.51.1415}
  {\path{doi:10.1103/PhysRevLett.51.1415}}.
\newline\urlprefix\url{http://link.aps.org/doi/10.1103/PhysRevLett.51.1415}

\bibitem{Sikivie1985}
P.~Sikivie, \href{http://link.aps.org/doi/10.1103/PhysRevD.32.2988}{Detection
  rates for ``invisible''-axion searches}, Phys. Rev. D 32 (1985) 2988--2991.
\newblock \href {http://dx.doi.org/10.1103/PhysRevD.32.2988}
  {\path{doi:10.1103/PhysRevD.32.2988}}.
\newline\urlprefix\url{http://link.aps.org/doi/10.1103/PhysRevD.32.2988}

\bibitem{ADMXaxions2010}
S.~J. Asztalos, G.~Carosi, C.~Hagmann, D.~Kinion, K.~van Bibber, M.~Hotz, L.~J.
  Rosenberg, G.~Rybka, J.~Hoskins, J.~Hwang, P.~Sikivie, D.~B. Tanner,
  R.~Bradley, J.~Clarke,
  \href{http://link.aps.org/doi/10.1103/PhysRevLett.104.041301}{Squid-based
  microwave cavity search for dark-matter axions}, Phys. Rev. Lett. 104 (2010)
  041301.
\newblock \href {http://dx.doi.org/10.1103/PhysRevLett.104.041301}
  {\path{doi:10.1103/PhysRevLett.104.041301}}.
\newline\urlprefix\url{http://link.aps.org/doi/10.1103/PhysRevLett.104.041301}

\bibitem{ADMX2011}
J.~Hoskins, J.~Hwang, C.~Martin, P.~Sikivie, N.~S. Sullivan, D.~B. Tanner,
  M.~Hotz, L.~J. Rosenberg, G.~Rybka, A.~Wagner, S.~J. Asztalos, G.~Carosi,
  C.~Hagmann, D.~Kinion, K.~van Bibber, R.~Bradley, J.~Clarke,
  \href{http://link.aps.org/doi/10.1103/PhysRevD.84.121302}{Search for
  nonvirialized axionic dark matter}, Phys. Rev. D 84 (2011) 121302.
\newblock \href {http://dx.doi.org/10.1103/PhysRevD.84.121302}
  {\path{doi:10.1103/PhysRevD.84.121302}}.
\newline\urlprefix\url{http://link.aps.org/doi/10.1103/PhysRevD.84.121302}

\bibitem{Sikivie1983}
L.~Abbott, P.~Sikivie,
  \href{http://www.sciencedirect.com/science/article/pii/037026938390638X}{A
  cosmological bound on the invisible axion}, Physics Letters B 120~(1–3)
  (1983) 133 -- 136.
\newblock \href
  {http://dx.doi.org/http://dx.doi.org/10.1016/0370-2693(83)90638-X}
  {\path{doi:http://dx.doi.org/10.1016/0370-2693(83)90638-X}}.
\newline\urlprefix\url{http://www.sciencedirect.com/science/article/pii/037026938390638X}

\bibitem{Preskill1983}
J.~Preskill, M.~B. Wise, F.~Wilczek,
  \href{http://www.sciencedirect.com/science/article/pii/0370269383906378}{Cosmology
  of the invisible axion}, Physics Letters B 120~(1) (1983) 127 -- 132.
\newblock \href
  {http://dx.doi.org/http://dx.doi.org/10.1016/0370-2693(83)90637-8}
  {\path{doi:http://dx.doi.org/10.1016/0370-2693(83)90637-8}}.
\newline\urlprefix\url{http://www.sciencedirect.com/science/article/pii/0370269383906378}

\bibitem{K79}
J.~E. Kim,
  \href{http://link.aps.org/doi/10.1103/PhysRevLett.43.103}{Weak-interaction
  singlet and strong $\mathrm{CP}$ invariance}, Phys. Rev. Lett. 43 (1979)
  103--107.
\newblock \href {http://dx.doi.org/10.1103/PhysRevLett.43.103}
  {\path{doi:10.1103/PhysRevLett.43.103}}.
\newline\urlprefix\url{http://link.aps.org/doi/10.1103/PhysRevLett.43.103}

\bibitem{Kim2010}
J.~E. Kim, G.~Carosi,
  \href{http://link.aps.org/doi/10.1103/RevModPhys.82.557}{Axions and the
  strong $cp$ problem}, Rev. Mod. Phys. 82 (2010) 557--601.
\newblock \href {http://dx.doi.org/10.1103/RevModPhys.82.557}
  {\path{doi:10.1103/RevModPhys.82.557}}.
\newline\urlprefix\url{http://link.aps.org/doi/10.1103/RevModPhys.82.557}

\bibitem{DFS81}
M.~Dine, W.~Fischler, M.~Srednicki,
  \href{http://www.sciencedirect.com/science/article/pii/0370269381905906}{A
  simple solution to the strong \{CP\} problem with a harmless axion}, Physics
  Letters B 104~(3) (1981) 199 -- 202.
\newblock \href
  {http://dx.doi.org/http://dx.doi.org/10.1016/0370-2693(81)90590-6}
  {\path{doi:http://dx.doi.org/10.1016/0370-2693(81)90590-6}}.
\newline\urlprefix\url{http://www.sciencedirect.com/science/article/pii/0370269381905906}

\bibitem{SVZ80}
M.~Shifman, A.~Vainshtein, V.~Zakharov,
  \href{http://www.sciencedirect.com/science/article/pii/0550321380902096}{Can
  confinement ensure natural \{CP\} invariance of strong interactions?},
  Nuclear Physics B 166~(3) (1980) 493 -- 506.
\newblock \href
  {http://dx.doi.org/http://dx.doi.org/10.1016/0550-3213(80)90209-6}
  {\path{doi:http://dx.doi.org/10.1016/0550-3213(80)90209-6}}.
\newline\urlprefix\url{http://www.sciencedirect.com/science/article/pii/0550321380902096}

\bibitem{Dine1983}
M.~Dine, W.~Fischler,
  \href{http://www.sciencedirect.com/science/article/pii/0370269383906391}{The
  not-so-harmless axion}, Physics Letters B 120~(1) (1983) 137 -- 141.
\newblock \href
  {http://dx.doi.org/http://dx.doi.org/10.1016/0370-2693(83)90639-1}
  {\path{doi:http://dx.doi.org/10.1016/0370-2693(83)90639-1}}.
\newline\urlprefix\url{http://www.sciencedirect.com/science/article/pii/0370269383906391}

\bibitem{SMASH}
G.~Ballesteros, J.~Redondo, A.~Ringwald, C.~Tamarit, {Unifying inflation with
  the axion, dark matter, baryogenesis and the seesaw mechanism}, Phys. Rev.
  Lett. 118 (2017) 071802.
\newblock \href {http://arxiv.org/abs/1608.05414} {\path{arXiv:1608.05414}},
  \href {http://dx.doi.org/10.1103/PhysRevLett.118.071802}
  {\path{doi:10.1103/PhysRevLett.118.071802}}.

\bibitem{Beck2013}
C.~Beck, \href{http://link.aps.org/doi/10.1103/PhysRevLett.111.231801}{Possible
  resonance effect of axionic dark matter in josephson junctions}, Phys. Rev.
  Lett. 111 (2013) 231801.
\newblock \href {http://dx.doi.org/10.1103/PhysRevLett.111.231801}
  {\path{doi:10.1103/PhysRevLett.111.231801}}.
\newline\urlprefix\url{http://link.aps.org/doi/10.1103/PhysRevLett.111.231801}

\bibitem{Beck2014}
C.~Beck, {Axion mass estimates from resonant Josephson junctions}, Phys. Dark
  Univ. 7-8 (2015) 6--11.
\newblock \href {http://arxiv.org/abs/1403.5676} {\path{arXiv:1403.5676}},
  \href {http://dx.doi.org/10.1016/j.dark.2015.03.002}
  {\path{doi:10.1016/j.dark.2015.03.002}}.

\bibitem{darin2001}
D.~S. Kinion, First results from a multiple-microwave-cavity search for
  dark-matter axions, Ph.D. thesis, UC Davis (2001).

\bibitem{Sikivie2014a}
P.~Sikivie, N.~Sullivan, D.~B. Tanner,
  \href{http://link.aps.org/doi/10.1103/PhysRevLett.112.131301}{Proposal for
  axion dark matter detection using an $lc$ circuit}, Phys. Rev. Lett. 112
  (2014) 131301.
\newblock \href {http://dx.doi.org/10.1103/PhysRevLett.112.131301}
  {\path{doi:10.1103/PhysRevLett.112.131301}}.
\newline\urlprefix\url{http://link.aps.org/doi/10.1103/PhysRevLett.112.131301}

\bibitem{McAllister:2016fux}
B.~T. McAllister, S.~R. Parker, M.~E. Tobar, {3D Lumped LC Resonators as Low
  Mass Axion Haloscopes}, Phys. Rev. D94~(4) (2016) 042001.
\newblock \href {http://arxiv.org/abs/1605.05427} {\path{arXiv:1605.05427}},
  \href {http://dx.doi.org/10.1103/PhysRevD.94.042001}
  {\path{doi:10.1103/PhysRevD.94.042001}}.

\bibitem{ABRACADABRA}
Y.~Kahn, B.~R. Safdi, J.~Thaler, {Broadband and Resonant Approaches to Axion
  Dark Matter Detection}, Phys. Rev. Lett. 117~(14) (2016) 141801.
\newblock \href {http://arxiv.org/abs/1602.01086} {\path{arXiv:1602.01086}},
  \href {http://dx.doi.org/10.1103/PhysRevLett.117.141801}
  {\path{doi:10.1103/PhysRevLett.117.141801}}.

\bibitem{MADMAX}
A.~Caldwell, G.~Dvali, B.~Majorovits, A.~Millar, G.~Raffelt, J.~Redondo,
  O.~Reimann, F.~Simon, F.~Steffen,
  \href{https://link.aps.org/doi/10.1103/PhysRevLett.118.091801}{Dielectric
  haloscopes: A new way to detect axion dark matter}, Phys. Rev. Lett. 118
  (2017) 091801.
\newblock \href {http://dx.doi.org/10.1103/PhysRevLett.118.091801}
  {\path{doi:10.1103/PhysRevLett.118.091801}}.
\newline\urlprefix\url{https://link.aps.org/doi/10.1103/PhysRevLett.118.091801}

\bibitem{ADMXHF2014}
T.~M. Shokair, J.~Root, K.~A. Van~Bibber, B.~Brubaker, Y.~V. Gurevich, S.~B.
  Cahn, S.~K. Lamoreaux, M.~A. Anil, K.~W. Lehnert, B.~K. Mitchell, A.~Reed,
  G.~Carosi, Future directions in the microwave cavity search for dark matter
  axions, International Journal of Modern Physics A 29~(19) (2014) 1443004.
\newblock \href {http://dx.doi.org/10.1142/S0217751X14430040}
  {\path{doi:10.1142/S0217751X14430040}}.

\bibitem{YaleAxion}
B.~M. Brubaker, L.~Zhong, Y.~V. Gurevich, S.~B. Cahn, S.~K. Lamoreaux,
  M.~Simanovskaia, J.~R. Root, S.~M. Lewis, S.~Al~Kenany, K.~M. Backes,
  I.~Urdinaran, N.~M. Rapidis, T.~M. Shokair, K.~A. van Bibber, D.~A. Palken,
  M.~Malnou, W.~F. Kindel, M.~A. Anil, K.~W. Lehnert, G.~Carosi,
  \href{https://link.aps.org/doi/10.1103/PhysRevLett.118.061302}{First results
  from a microwave cavity axion search at $24\text{ }\text{
  }\ensuremath{\mu}\mathrm{eV}$}, Phys. Rev. Lett. 118 (2017) 061302.
\newblock \href {http://dx.doi.org/10.1103/PhysRevLett.118.061302}
  {\path{doi:10.1103/PhysRevLett.118.061302}}.
\newline\urlprefix\url{https://link.aps.org/doi/10.1103/PhysRevLett.118.061302}

\bibitem{CULTASK}
W.~Chung, {CULTASK, The Coldest Axion Experiment at CAPP/IBS in Korea}, PoS
  CORFU2015 (2016) 047.

\bibitem{Jaeckel2013}
J.~Jaeckel, J.~Redondo,
  \href{http://link.aps.org/doi/10.1103/PhysRevD.88.115002}{Resonant to
  broadband searches for cold dark matter consisting of weakly interacting slim
  particles}, Phys. Rev. D 88 (2013) 115002.
\newblock \href {http://dx.doi.org/10.1103/PhysRevD.88.115002}
  {\path{doi:10.1103/PhysRevD.88.115002}}.
\newline\urlprefix\url{http://link.aps.org/doi/10.1103/PhysRevD.88.115002}

\bibitem{SpringsPaper}
M.~Goryachev, B.~T. Mcallister, M.~E. Tobar, {Axion Detection with Cavity
  Arrays}\href {http://arxiv.org/abs/1703.07207} {\path{arXiv:1703.07207}}.

\bibitem{Daw:1998jm}
E.~J. Daw, \href{http://wwwlib.umi.com/dissertations/fullcit?334417}{{A search
  for halo axions}}, Ph.D. thesis, MIT (1998).
\newline\urlprefix\url{http://wwwlib.umi.com/dissertations/fullcit?334417}

\bibitem{McAllisterFormFactor}
B.~T. McAllister, S.~R. Parker, M.~E. Tobar, {Axion Dark Matter Coupling to
  Resonant Photons via Magnetic Field}, Phys. Rev. Lett. 116~(16) (2016)
  161804, [Erratum: Phys. Rev. Lett.117,no.15,159901(2016)].
\newblock \href {http://arxiv.org/abs/1607.01928} {\path{arXiv:1607.01928}},
  \href {http://dx.doi.org/10.1103/PhysRevLett.117.159901,
  10.1103/PhysRevLett.116.161804} {\path{doi:10.1103/PhysRevLett.117.159901,
  10.1103/PhysRevLett.116.161804}}.

\bibitem{Stern:2015kzo}
I.~Stern, A.~A. Chisholm, J.~Hoskins, P.~Sikivie, N.~S. Sullivan, D.~B. Tanner,
  G.~Carosi, K.~van Bibber, {Cavity design for high-frequency axion dark matter
  detectors}, Rev. Sci. Instrum. 86~(12) (2015) 123305.
\newblock \href {http://arxiv.org/abs/1603.06990} {\path{arXiv:1603.06990}},
  \href {http://dx.doi.org/10.1063/1.4938164} {\path{doi:10.1063/1.4938164}}.

\bibitem{ORGANPatras2016}
B.~T. McAllister, S.~R. Parker, E.~N. Ivanov, M.~E. Tobar,
  \href{http://inspirehep.net/record/1500189/files/arXiv:1611.08082.pdf}{{High
  and low mass Axion Haloscopes at UWA}}, in: {12th Patras Workshop on Axions,
  WIMPs and WISPs (AXION-WIMP 2016) Jeju Island, South Korea, June 20-24,
  2016}, 2016.
\newblock \href {http://arxiv.org/abs/1611.08082} {\path{arXiv:1611.08082}}.
\newline\urlprefix\url{http://inspirehep.net/record/1500189/files/arXiv:1611.08082.pdf}

\bibitem{DielectricSupermode}
B.~T. McAllister, G.~Flower, L.~E. Tobar, M.~E. Tobar, {Tunable Super-Mode
  Dielectric Resonators for Axion Haloscopes}\href
  {http://arxiv.org/abs/1705.06028} {\path{arXiv:1705.06028}}.

\bibitem{CAST}
V.~Anastassopoulos, S.~Aune, K.~Barth, A.~Belov, H.~Bräuninger, G.~Cantatore,
  J.~M. Carmona, J.~F. Castel, S.~A. Cetin, F.~Christensen, J.~I. Collar,
  T.~Dafni, M.~Davenport, T.~A. Decker, A.~Dermenev, K.~Desch,
  C.~Eleftheriadis, G.~Fanourakis, E.~Ferrer-Ribas, H.~Fischer, J.~A. García,
  A.~Gardikiotis, J.~G. Garza, E.~N. Gazis, T.~Geralis, I.~Giomataris,
  S.~Gninenko, C.~J. Hailey, M.~D. Hasinoff, D.~H.~H. Hoffmann, F.~J. Iguaz,
  I.~G. Irastorza, A.~Jakobsen, J.~Jacoby, K.~Jakovčić, J.~Kaminski,
  M.~Karuza, N.~Kralj, M.~Krčmar, S.~Kostoglou, C.~Krieger, B.~Lakić, J.~M.
  Laurent, A.~Liolios, A.~Ljubičić, G.~Luzón, M.~Maroudas, L.~Miceli,
  S.~Neff, I.~Ortega, T.~Papaevangelou, K.~Paraschou, M.~J. Pivovaroff,
  G.~Raffelt, M.~Rosu, J.~Ruz, E.~R. Chóliz, I.~Savvidis, S.~Schmidt, Y.~K.
  Semertzidis, S.~K. Solanki, L.~Stewart, T.~Vafeiadis, J.~K. Vogel, S.~C.
  Yildiz, K.~Zioutas, \href{http:https://dx.doi.org/10.1038/nphys4109}{New cast
  limit on the axion-photon interaction}, Nature Physics\href
  {http://dx.doi.org/10.1038/nphys4109} {\path{doi:10.1038/nphys4109}}.
\newline\urlprefix\url{http:https://dx.doi.org/10.1038/nphys4109}

\bibitem{XSWisp}
B.~T. McAllister, S.~R. Parker, E.~N. Ivanov, M.~E. Tobar, {Cross-correlation
  measurement techniques for cavity-based axion and weakly interacting slim
  particle searches}\href {http://arxiv.org/abs/1510.05775}
  {\path{arXiv:1510.05775}}.

\bibitem{cdmdensity2014}
P.~R. Kafle, S.~Sharma, G.~F. Lewis, J.~Bland-Hawthorn,
  \href{http://stacks.iop.org/0004-637X/794/i=1/a=59}{On the shoulders of
  giants: Properties of the stellar halo and the milky way mass distribution},
  The Astrophysical Journal 794~(1) (2014) 59.
\newline\urlprefix\url{http://stacks.iop.org/0004-637X/794/i=1/a=59}

\bibitem{RBF1}
S.~De~Panfilis, A.~C. Melissinos, B.~E. Moskowitz, J.~T. Rogers, Y.~K.
  Semertzidis, W.~Wuensch, H.~J. Halama, A.~G. Prodell, W.~B. Fowler, F.~A.
  Nezrick, {Limits on the Abundance and Coupling of Cosmic Axions at
  4.5-Microev < m(a) < 5.0-Microev}, Phys. Rev. Lett. 59 (1987) 839.
\newblock \href {http://dx.doi.org/10.1103/PhysRevLett.59.839}
  {\path{doi:10.1103/PhysRevLett.59.839}}.

\bibitem{RBF2}
W.~Wuensch, S.~De~Panfilis-Wuensch, Y.~K. Semertzidis, J.~T. Rogers, A.~C.
  Melissinos, H.~J. Halama, B.~E. Moskowitz, A.~G. Prodell, W.~B. Fowler, F.~A.
  Nezrick, {Results of a Laboratory Search for Cosmic Axions and Other Weakly
  Coupled Light Particles}, Phys. Rev. D40 (1989) 3153.
\newblock \href {http://dx.doi.org/10.1103/PhysRevD.40.3153}
  {\path{doi:10.1103/PhysRevD.40.3153}}.

\bibitem{UFaxions}
C.~Hagmann, P.~Sikivie, N.~S. Sullivan, D.~B. Tanner, {Results from a search
  for cosmic axions}, Phys. Rev. D42 (1990) 1297--1300.
\newblock \href {http://dx.doi.org/10.1103/PhysRevD.42.1297}
  {\path{doi:10.1103/PhysRevD.42.1297}}.

\bibitem{ANULIGO}
J.~Aasi, et~al., {Enhancing the sensitivity of the LIGO gravitational wave
  detector by using squeezed states of light}, Nature Photon. 7 (2013)
  613--619.
\newblock \href {http://arxiv.org/abs/1310.0383} {\path{arXiv:1310.0383}},
  \href {http://dx.doi.org/10.1038/nphoton.2013.177}
  {\path{doi:10.1038/nphoton.2013.177}}.

\bibitem{Nobel2012}
S.~Haroche, \href{https://link.aps.org/doi/10.1103/RevModPhys.85.1083}{Nobel
  lecture: Controlling photons in a box and exploring the quantum to classical
  boundary}, Rev. Mod. Phys. 85 (2013) 1083--1102.
\newblock \href {http://dx.doi.org/10.1103/RevModPhys.85.1083}
  {\path{doi:10.1103/RevModPhys.85.1083}}.
\newline\urlprefix\url{https://link.aps.org/doi/10.1103/RevModPhys.85.1083}

\bibitem{AxionQND3}
H.~Zheng, M.~Silveri, R.~T. Brierley, S.~M. Girvin, K.~W. Lehnert,
  {Accelerating dark-matter axion searches with quantum measurement
  technology}\href {http://arxiv.org/abs/1607.02529} {\path{arXiv:1607.02529}}.

\bibitem{AxionQND1}
S.~K. Lamoreaux, K.~A. van Bibber, K.~W. Lehnert, G.~Carosi,
  \href{https://link.aps.org/doi/10.1103/PhysRevD.88.035020}{Analysis of
  single-photon and linear amplifier detectors for microwave cavity dark matter
  axion searches}, Phys. Rev. D 88 (2013) 035020.
\newblock \href {http://dx.doi.org/10.1103/PhysRevD.88.035020}
  {\path{doi:10.1103/PhysRevD.88.035020}}.
\newline\urlprefix\url{https://link.aps.org/doi/10.1103/PhysRevD.88.035020}

\bibitem{AxionQND2}
A.~Kitagawa, K.~Yamamoto, S.~Matsuki, {Quantum analysis of Rydberg atom cavity
  detector for dark matter axion search}\href
  {http://arxiv.org/abs/hep-ph/9908445} {\path{arXiv:hep-ph/9908445}}.

\bibitem{Orpheus}
G.~Rybka, A.~Wagner, A.~Brill, K.~Ramos, R.~Percival, K.~Patel, {Search for
  dark matter axions with the Orpheus experiment}, Phys. Rev. D91~(1) (2015)
  011701.
\newblock \href {http://arxiv.org/abs/1403.3121} {\path{arXiv:1403.3121}},
  \href {http://dx.doi.org/10.1103/PhysRevD.91.011701}
  {\path{doi:10.1103/PhysRevD.91.011701}}.

\end{thebibliography}

\end{document}